\documentclass[aps,preprintnumbers,superscriptaddress,showpacs]{revtex4}
\usepackage{epsfig}
\usepackage{psfrag}
\usepackage{amsfonts}
\usepackage{graphicx}
\usepackage{dcolumn}
\usepackage{bm}

\begin{document}

\title{Holographic Schwinger effect in a soft wall AdS/QCD
model}

\author{Yue Ding}
\affiliation{School of Mathematics and Physics, China University
of Geosciences, Wuhan 430074, China}

\author{Zi-qiang Zhang}
\email{zhangzq@cug.edu.cn} \affiliation{School of Mathematics and
Physics, China University of Geosciences, Wuhan 430074, China}

\begin{abstract}
We perform the potential analysis for the holographic Schwinger
effect in a deformed $AdS_5$ model with conformal invariance
broken by a background dilaton. We evaluate the static potential
by analyzing the classical action of a string attaching the
rectangular Wilson loop on a probe D3 brane sitting at an
intermediate position in the bulk AdS space. We observe that the
inclusion of chemical potential tends to enhance the production
rate, reverse to the effect of confining scale. Also, we calculate
the critical electric field by Dirac-Born-Infeld (DBI) action.

\end{abstract}
\pacs{11.25.Tq, 11.15.Tk, 11.25-w}

\maketitle
\section{Introduction}
Schwinger effect is an interesting phenomenon in quantum
electrodynamics (QED): virtual electron-position pairs can be
materialized and become real particles due to the presence of a
strong electric field. The production rate $\Gamma$ (per unit time
and unit volume) was first calculated by Schwinger for
weak-coupling and weak-field in 1951 \cite{JS}
\begin{equation}
\Gamma\sim exp\Big({\frac{-\pi m^2}{eE}}\Big),
\end{equation}
where $E$, $m$ and $e$ are the external electric field, electron
mass and elementary electric charge, respectively. In this case,
there is no critical field trivially. Thirty-one years later,
Affleck et.al generalized it to the case for arbitrary-coupling
and weak-field \cite{IK}
\begin{equation}
\Gamma\sim exp\Big({\frac{-\pi m^2}{eE}+\frac{e^2}{4}}\Big),
\end{equation}
in this case, the exponential suppression vanishes when $E$
reaches $E_c=(4\pi/e^3)m^2\simeq137m^2/e$. Obviously, the critical
field $E_c$ does not satisfy the weak-field condition, i.e.,
$eE\ll m^2$. Thus, it seems that one could not find out $E_c$
under the weak-field condition. One step further, one doesn't know
whether the catastrophic decay really occurs or not.

Actually, the Schwinger effect is not unique to QED, but a
universal aspect of quantum field theories (QFTs) coupled to an
U(1) gauge field. However, it remains difficult to study this
effect in a QCD-like or confining theory using QFTs since the
(original) Schwinger effect must be non-perturbative. Fortunately,
the AdS/CFT correspondence
\cite{Maldacena:1997re,Gubser:1998bc,MadalcenaReview} may provide
an alternative way. In 2011, Semenoff and Zarembo proposed
\cite{GW} a holographic set-up to study the Schwinger effect in
the higgsed $\mathcal N=4$ supersymmetric Yang-Mills theory (SYM).
They found that at large $N$ and large 't Hooft coupling $\lambda$
\begin{equation}
\Gamma\sim
exp\Big[-\frac{\sqrt{\lambda}}{2}\Big(\sqrt{\frac{E_c}{E}}-\sqrt{\frac{E}{E_c}}\Big)^2\Big],
\qquad E_c=\frac{2\pi m^2}{\sqrt{\lambda}},\label{gama}
\end{equation}
interestingly, the value of $E_c$ coincides with the one obtained
from the DBI action \cite{YS0}. Subsequently, Sato and Yoshida
argued that \cite{YS} the Schwinger effect can be studied by
potential analysis. Specifically, the pair production can be
estimated by a static potential, consisting of static mass
energies, an electric potential from an external electric-field,
and the Coulomb potential between a particle-antiparticle pair.
The shapes of the potential depend on the external field $E$ (see
fig.1 ). When $E<E_c$, the potential barrier is present and the
Schwinger effect could occur as a tunneling process. As $E$
increases, the barrier decreases and gradually disappears at
$E=E_c$. When $E>E_c$, the vacuum becomes catastrophically
unstable. Further studies of the Schwinger effect in this
direction can be found, e.g., in
\cite{YS1,YS2,SCH,KB,MG,ZL,ZQ,ZQ1,LS,WF,ZR}. On the other hand,
the holographic Schwinger effect has been investigated from the
imaginary part of a probe brane action \cite{KHA,KHA1,XW,KG}. For
a recent review on this topic, see \cite{DK}.

\begin{figure}
\centering
\includegraphics[width=10cm]{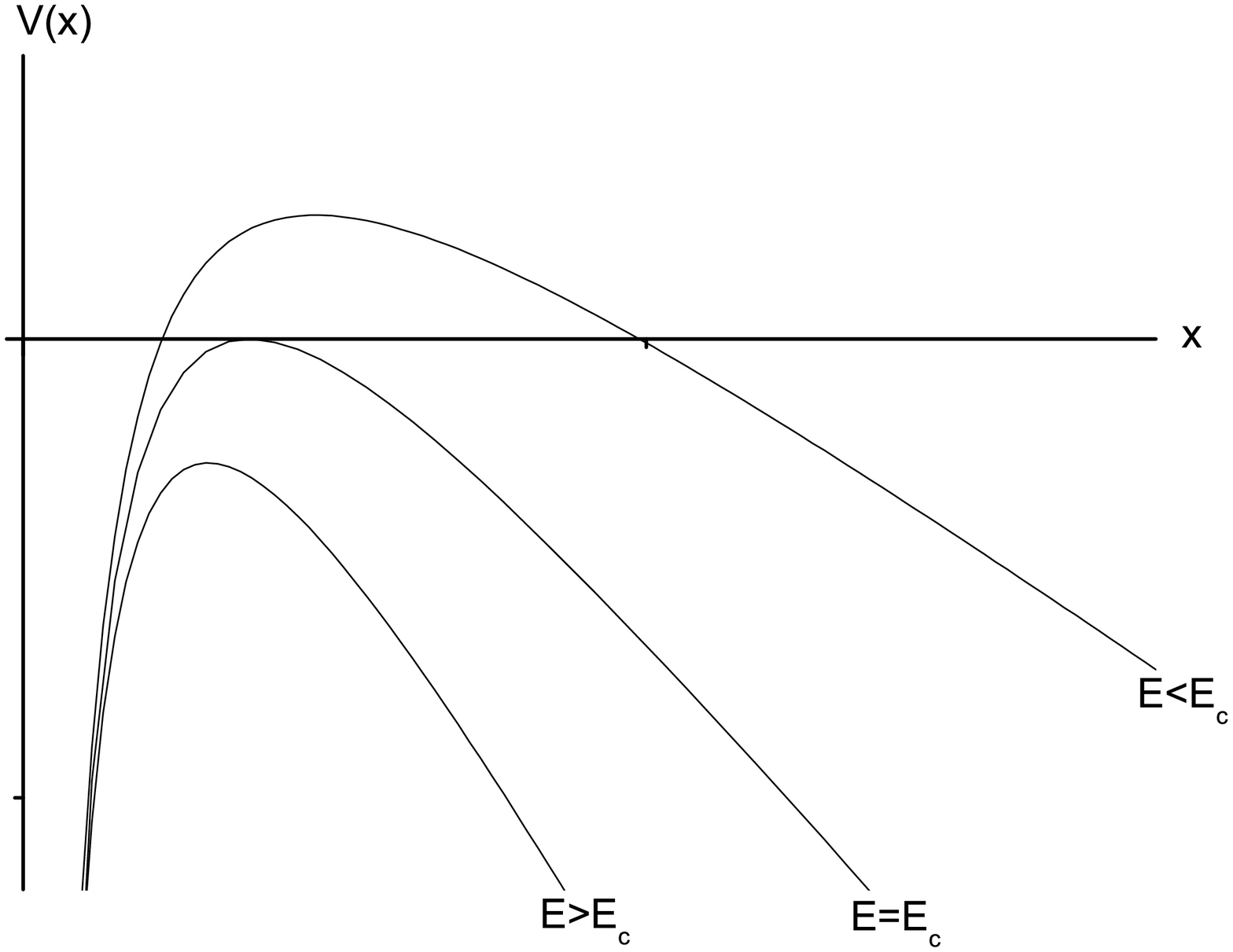}
\caption{$V(x)$ versus $x$ with $V(x)=2m-eEx-\frac{\alpha_s}{x}$,
where $\alpha_s$ denotes the fine-structure constant.}
\end{figure}

Here we present an alternative holographic approach to study the
Schiwinger effect using potential analysis. The motivation is that
holographic QCD models, like hard wall \cite{H1,H2}, soft wall
\cite{AKE} and some improved AdS/QCD models
\cite{JP,AST,DL,DL1,SH,SH1,RRO} have achieved considerable success
in describing various aspects of hadron physics. In particular, we
will adopt the SW$_{T,\mu}$ model \cite{PCO} which is defined by
the AdS with a charged black hole to describe finite temperature
and density multiplied by a warp factor to generate confinement.
It turns out that such a model can provide a good phenomenological
description of quark-antiquark interaction. Also, the resulting
deconfinement line in $\mu-T$ plane is similar to that obtained by
lattice and effective models of QCD (for further studies of models
of this type, see \cite{CPA,PCO1,PCO2,YH,XCH,zq}). Motivated by
this, in this paper we study the Schwinger effect in the
SW$_{T,\mu}$ model. Specifically, we want to understand how the
Schwinger effect is affected by chemical potential and confining
scale. Also, this work could be considered as an extension of
\cite{YS} to the case with chemical potential and confining scale.

The outline of the paper is as follows. In the next section, we
briefly review the SW$_{T,\mu}$ model given in \cite{PCO}. In
section 3, we perform the potential analysis for the Schwinger
effect in the SW$_{T,\mu}$ model and investigate how chemical
potential and confining scale affect the production rate. Also, we
calculate the critical field from DBI action. Finally, we conclude
our results in section 4.

\section{Setup}
This section is devoted to a short introduction of the
SW$_{T,\mu}$ model proposed in \cite{PCO}. The metric of the model
in the string frame takes the form
\begin{equation}
ds^2=\frac{R^2}{z^2}h(z)(-f(z)dt^2+d\vec{x}^2+\frac{dz^2}{f(z)}),\label{metric0}
\end{equation}
with
\begin{equation}
f(z)=1-(1+Q^2)(\frac{z}{z_h})^4+Q^2(\frac{z}{z_h})^6, \qquad
h(z)=e^{c^2z^2},
\end{equation}
where $R$ is the AdS radius. $Q$ represents the charge of black
hole. $z$ denotes the fifth coordinate with $z=z_h$ the horizon,
defined by $f(z_h)=0$. The warp factor $h(z)$, characterizing the
soft wall model, distorts the metric and brings the confining
scale $c$ (see \cite{JP} for a anatlytical way to introduce the
warp factor by potential reconstruction approach).

The temperature of the black hole is
\begin{equation}
T=\frac{1}{\pi z_h}(1-\frac{Q^2}{2}), \qquad 0\leq
Q\leq\sqrt{2}.\label{T}
\end{equation}

The chemical potential is
\begin{equation}
\mu=\sqrt{3}Q/z_h.
\end{equation}

Note that for $Q=0$, the SW$_{T,\mu}$ model reduces to the Andreev
model \cite{OA}. For $c=0$, it becomes the AdS-Reissner Nordstrom
black hole \cite{CV:1999,DT:2006}. For $Q=c=0$, it returns to AdS
black hole.

\section{Potential analysis in (holographic) Schwinger effect}
In this section, we follow the argument in \cite{YS} to study the
behavior of the Schwinger effect in the SW$_{T,\mu}$ model. Since
the calculations of \cite{YS} were performed using the radial
coordinate $r=R^2/z$. For contrast, we use coordinate $r$ as well.

The Nambu-Goto action is
\begin{equation}
S=T_F\int d\tau d\sigma\mathcal L=T_F\int d\tau d\sigma\sqrt{g},
\qquad  T_F=\frac{1}{2\pi\alpha^\prime},\label{S}
\end{equation}
where $\alpha^\prime$ is related to $\lambda$ by
$\frac{R^2}{\alpha^\prime}=\sqrt{\lambda}$. $g$ represents the
determinant of the induced metric
\begin{equation}
g_{\alpha\beta}=g_{\mu\nu}\frac{\partial
X^\mu}{\partial\sigma^\alpha} \frac{\partial
X^\nu}{\partial\sigma^\beta},
\end{equation}
with $g_{\mu\nu}$ the metric and $X^\mu$ the target space
coordinate.

Supposing the pair axis is aligned in one direction, e.g., $x_1$
direction,
\begin{equation}
t=\tau, \qquad x_1=\sigma, \qquad x_2=0,\qquad x_3=0, \qquad
r=r(\sigma). \label{par}
\end{equation}

Under this ansatz, the induced metric reads
\begin{equation} g_{00}=\frac{r^2h(r)f(r)}{R^2}, \qquad g_{01}=g_{10}=0, \qquad
g_{11}=\frac{r^2h(r)}{R^2}+\frac{R^2h(r)}{r^2f(r)}(\frac{dr}{d\sigma})^2,
\end{equation}
then the Lagrangian density becomes
\begin{equation}
\mathcal L=\sqrt{M(r)+N(r)(\frac{dr}{d\sigma})^2},\label{L}
\end{equation}
with
\begin{equation}
M(r)=\frac{r^4h^2(r)f(r)}{R^4},\qquad N(r)=h^2(r).
\end{equation}

As $\mathcal L$ does not depend on $\sigma$ explicitly, the
Hamiltonian is conserved,
\begin{equation}
\mathcal L-\frac{\partial\mathcal
L}{\partial(\frac{dr}{d\sigma})}(\frac{dr}{d\sigma})=Constant.
\end{equation}

Imposing the boundary condition at the tip of the minimal surface,
\begin{equation}
\frac{dr}{d\sigma}=0,\qquad  r=r_c\qquad (r_t<r_c<r_0)\label{con},
\end{equation}
one gets
\begin{equation}
\frac{dr}{d\sigma}=\sqrt{\frac{M^2(r)-M(r)M(r_c)}{M(r_c)N(r)}}\label{dotr},
\end{equation}
with $M(r_c)=M(r)|_{r=r_c}$. Here $r=r_t$ is the horizon. $r=r_0$
is an intermediate position, which can yield a finite mass
\cite{GW}. The configuration of the string world-sheet is depicted
in fig.2.
\begin{figure}
\centering
\includegraphics[width=10cm]{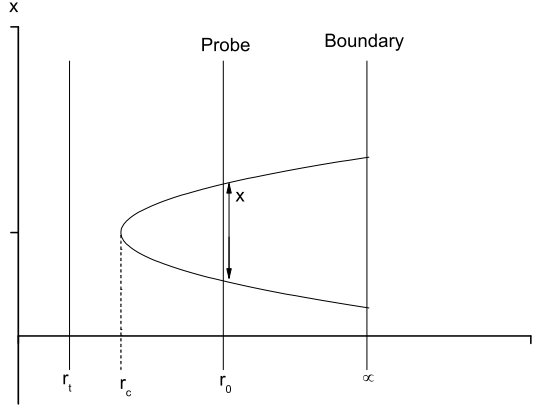}
\caption{The configuration of the string world-sheet.}
\end{figure}

Integrating (\ref{dotr}) with the boundary condition (\ref{con}),
the inter-distance of the particle pair is obtained
\begin{equation}
x=2\int_{r_c}^{r_0}\frac{d\sigma}{dr}dr=2\int_{r_c}^{r_0}dr\sqrt{\frac{M(r_c)N(r)}{M^2(r)-M(r)M(r_c)}}\label{xx}.
\end{equation}

On the other hand, plugging (\ref{L}) into (\ref{S}), the sum of
Coulomb potential and static energy is given by
\begin{equation}
V_{CP+E}=2T_F\int_{r_c}^{r_0}dr\sqrt{\frac{M(r)N(r)}{M(r)-M(r_c)}}.\label{en}
\end{equation}

The next task is to calculate the critical field. The DBI action
is
\begin{equation}
S_{DBI}=-T_{D3}\int
d^4x\sqrt{-det(G_{\mu\nu}+\mathcal{F}_{\mu\nu})}\label{dbi},
\end{equation}
with
\begin{equation}
T_{D3}=\frac{1}{g_s(2\pi)^3\alpha^{\prime^2}}, \qquad
\mathcal{F}_{\mu\nu}=2\pi\alpha^\prime F_{\mu\nu},
\end{equation}
where $T_{D3}$ is the D3-brane tension.

The induced metric is
\begin{equation}
G_{00}=-\frac{r^2h(r)f(r)}{R^2}, \qquad G_{11}=
G_{22}=G_{33}=\frac{r^2h(r)}{R^2}.
\end{equation}

Supposing the electric field is turned on along $x_1$ direction
\cite{YS}, then
\begin{equation}
G_{\mu\nu}+\mathcal{F}_{\mu\nu}=\left(
\begin{array}{cccc}
-\frac{r^2h(r)f(r)}{R^2} & 2\pi\alpha^\prime E & 0 & 0\\
 -2\pi\alpha^\prime E & \frac{r^2h(r)}{R^2} & 0 & 0 \\
 0 & 0 & \frac{r^2h(r)}{R^2} & 0\\
0 & 0 & 0 & \frac{r^2h(r)}{R^2}
\end{array}
\right),
\end{equation}
results in
\begin{equation}
det(G_{\mu\nu}+\mathcal{F}_{\mu\nu})=\frac{r^4h^2(r)}{R^4}[(2\pi\alpha^\prime)^2E^2-\frac{r^4h^2(r)f(r)}{R^4}].\label{det}
\end{equation}

Substituting (\ref{det}) into (\ref{dbi}) and making the probe
D3-brane located at $r=r_0$, one finds
\begin{equation}
S_{DBI}=-T_{D3}\frac{r_0^2h(r_0)}{R^2}\int d^4x
\sqrt{\frac{r_0^4h^2(r_0)f(r_0)}{R^4}-(2\pi\alpha^\prime)^2E^2}\label{dbi1},
\end{equation}
with $f(r_0)=f(r)|_{r=r_0}$, $h(r_0)=h(r)|_{r=r_0}$.

To avoid (\ref{dbi1}) being ill-defined, one gets
\begin{equation}
\frac{r_0^4h^2(r_0)f(r_0)}{R^4}-(2\pi\alpha^\prime)^2E^2\geq0,\label{ec}
\end{equation}
yielding
\begin{equation}
E\leq T_F\frac{r_0^2h(r_0)}{R^2}\sqrt{f(r_0)}.
\end{equation}

As a result, the critical field is
\begin{equation}
E_c=T_F\frac{r_0^2h(r_0)}{R^2}\sqrt{f(r_0)},\label{ec1}
\end{equation}
one can see that $E_c$ depends on $T$, $\mu$ and $c$.

Next, we calculate the total potential. For the sake of notation
simplicity, we introduce the following dimensionless parameters
\begin{equation}
\alpha\equiv\frac{E}{E_c}, \qquad y\equiv\frac{r}{r_c},\qquad
a\equiv\frac{r_c}{r_0},\qquad b\equiv\frac{r_t}{r_0}. \label{afa}
\end{equation}

Given that, the total potential reads
\begin{eqnarray}
V_{tot}(x)&=&V_{CP+E}-Ex\nonumber\\&=&2ar_0T_F\int_1^{1/a}dy\sqrt{\frac{A(y)B(y)}{A(y)-A(y_c)}}\nonumber\\&-&
2ar_0T_F\alpha
\frac{r_0^2h(y_0)}{R^2}\sqrt{f(y_0)}\int_1^{1/a}dy\sqrt{\frac{A(y_c)B(y)}{A^2(y)-A(y)A(y_c)}},
\label{V}
\end{eqnarray}
with
\begin{eqnarray}
A(y)&=&\frac{(ar_0y)^4h^2(y)f(y)}{R^4},\qquad
A(y_c)=\frac{(ar_0)^4h^2(y_c)f(y_c)}{R^4},\qquad B(y)=h^2(y)
,\nonumber\\
h(y)&=&e^{\frac{c^2R^4}{(ar_0y)^2}},\qquad
f(y)=1-(1+\frac{\mu^2R^4}{3r_t^2})(\frac{b}{ay})^4+\frac{\mu^2R^4}{3r_t^2}(\frac{b}{ay})^6, \nonumber\\
h(y_c)&=&e^{\frac{c^2R^4}{(ar_0)^2}},\qquad
f(y_c)=1-(1+\frac{\mu^2R^4}{3r_t^2})(\frac{b}{a})^4+\frac{\mu^2R^4}{3r_t^2}(\frac{b}{a})^6,\nonumber\\
h(y_0)&=&e^{\frac{c^2R^4}{r_0^2}},\qquad
f(y_0)=1-(1+\frac{\mu^2R^4}{3r_t^2})b^4+\frac{\mu^2R^4}{3r_t^2}b^6,
\end{eqnarray}
we have checked that by taking $c=\mu=0$ in (\ref{V}), the result
of $\mathcal{N}=4$ SYM \cite{YS} is regained.

\begin{figure}
\centering
\includegraphics[width=10cm]{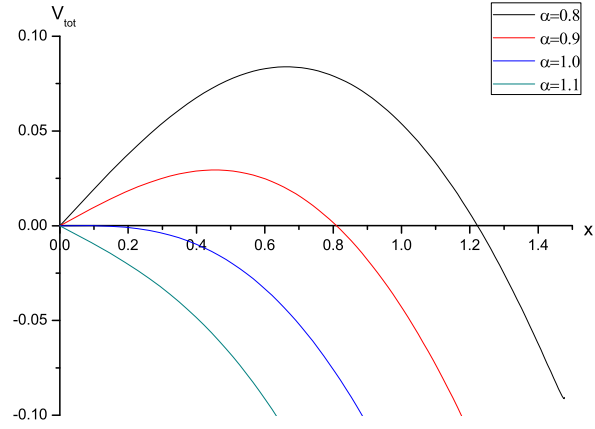}
\caption{$V_{tot}(x)$ versus $x$ with $\mu/T=1$, $c/T=0.1$. In the
plots from top to bottom $\alpha=0.8, 0.9, 1.0, 1.1$,
respectively.}
\end{figure}

Before going further, we discuss the value of $c$. In this work we
tend to study the behavior of the holographic Schwinger effect in
a class of models parametrized by $c$. To that end, we make $c$
dimensionless by normalizing it at fixed temperatures and express
other quantities, e.g, $\mu$, in units of it. In \cite{HLL}, the
authors found that the range of $0\leq c/T\leq2.5$ is most
relevant for a comparison with QCD. We use that range.

In fig.3, we plot $V_{tot}(x)$ as a function of $x$ for $\mu/T=1$
and $c/T=0.1$ (other cases with different values of $\mu/T$ and
$c/T$ have similar picture), where we have set $b=0.5$ and
$T_Fr_0=R^2/r_0=1$, as follows from \cite{YS}. From these figures,
one can see that there exists a critical electric field at
$\alpha=1$ ($E=E_c$), and for $\alpha<1$ ($E<E_c$), the potential
barrier is present, in agreement with \cite{YS}.

To see how chemical potential modifies the Schwinger effect, we
plot $V_{tot}(x)$ versus $x$ with fixed $c/T$ for different values
of $\mu/T$ in fig.4. The left panel is for $c/T=0.1$ and the right
$c/T=2.5$. In both panels from top to bottom $\mu/T=0, 1, 5$,
respectively. One can see that at fixed $c/T$, as $\mu/T$
increases, the height and width of the potential barrier both
decrease. As we know, the higher or the wider the potential
barrier, the harder the produced pairs escape to infinity. Thus,
one concludes that the inclusion of chemical potential decrease
the potential barrier thus enhancing the Schwinger effect, in
accordance with the findings of \cite{ZL}.

\begin{figure}
\centering
\includegraphics[width=8cm]{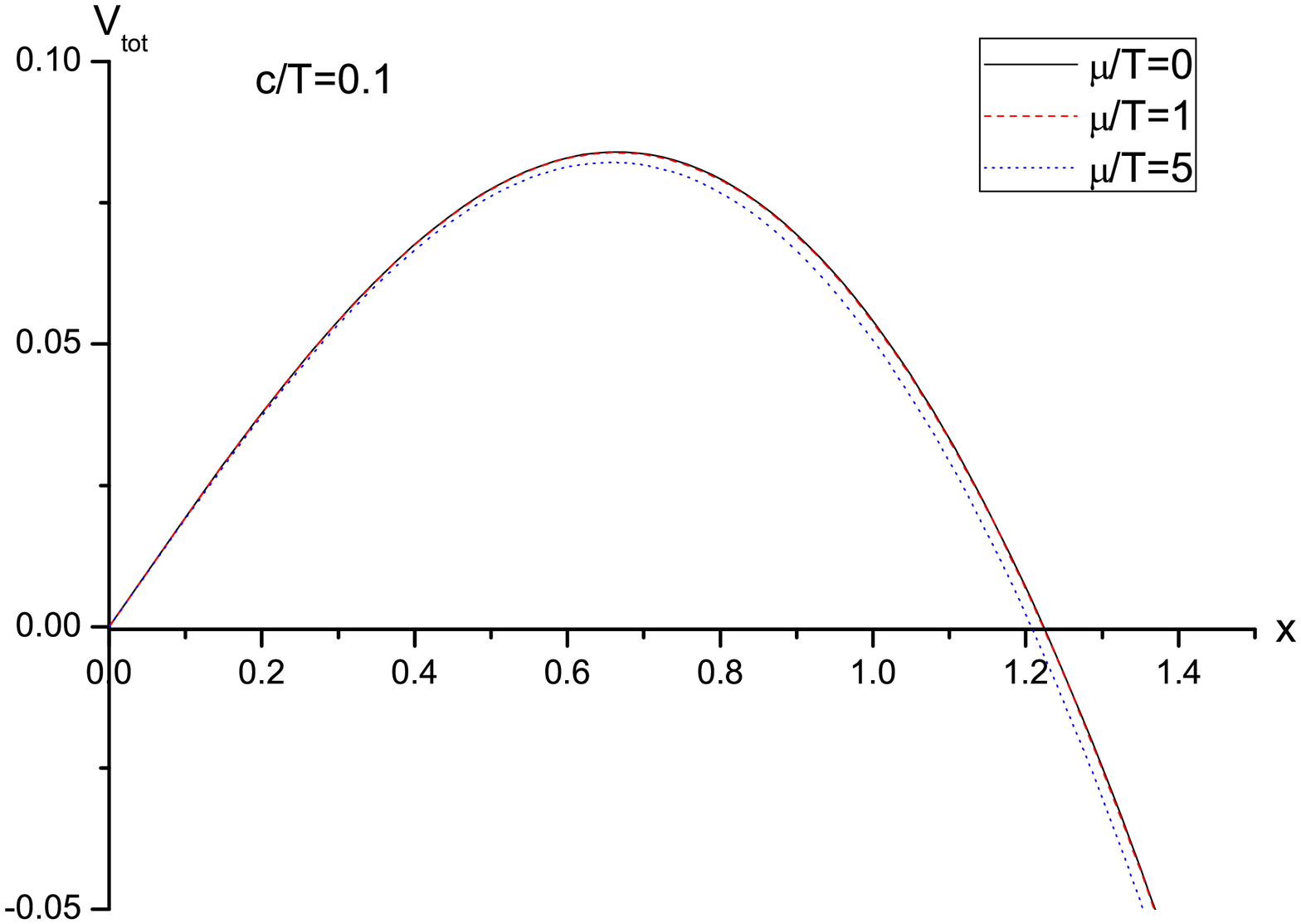}
\includegraphics[width=8cm]{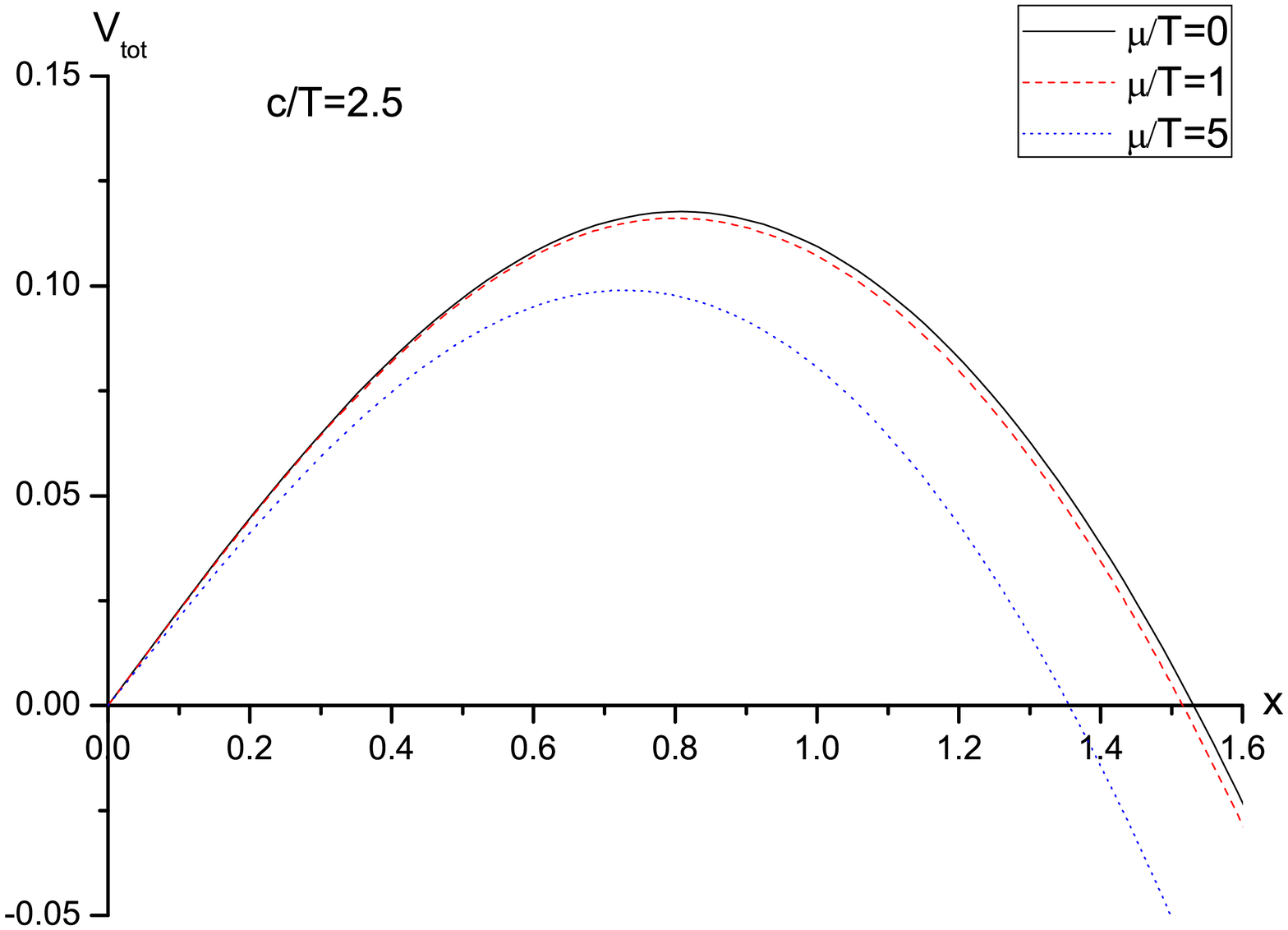}
\caption{$V_{tot}(x)$ versus $x$ with $\alpha=0.8$ and fixed $c/T$
for different values of $\mu/T$. In both plots from top to bottom
$\mu/T=0, 1, 5$, respectively.}
\end{figure}

\begin{figure}
\centering
\includegraphics[width=8cm]{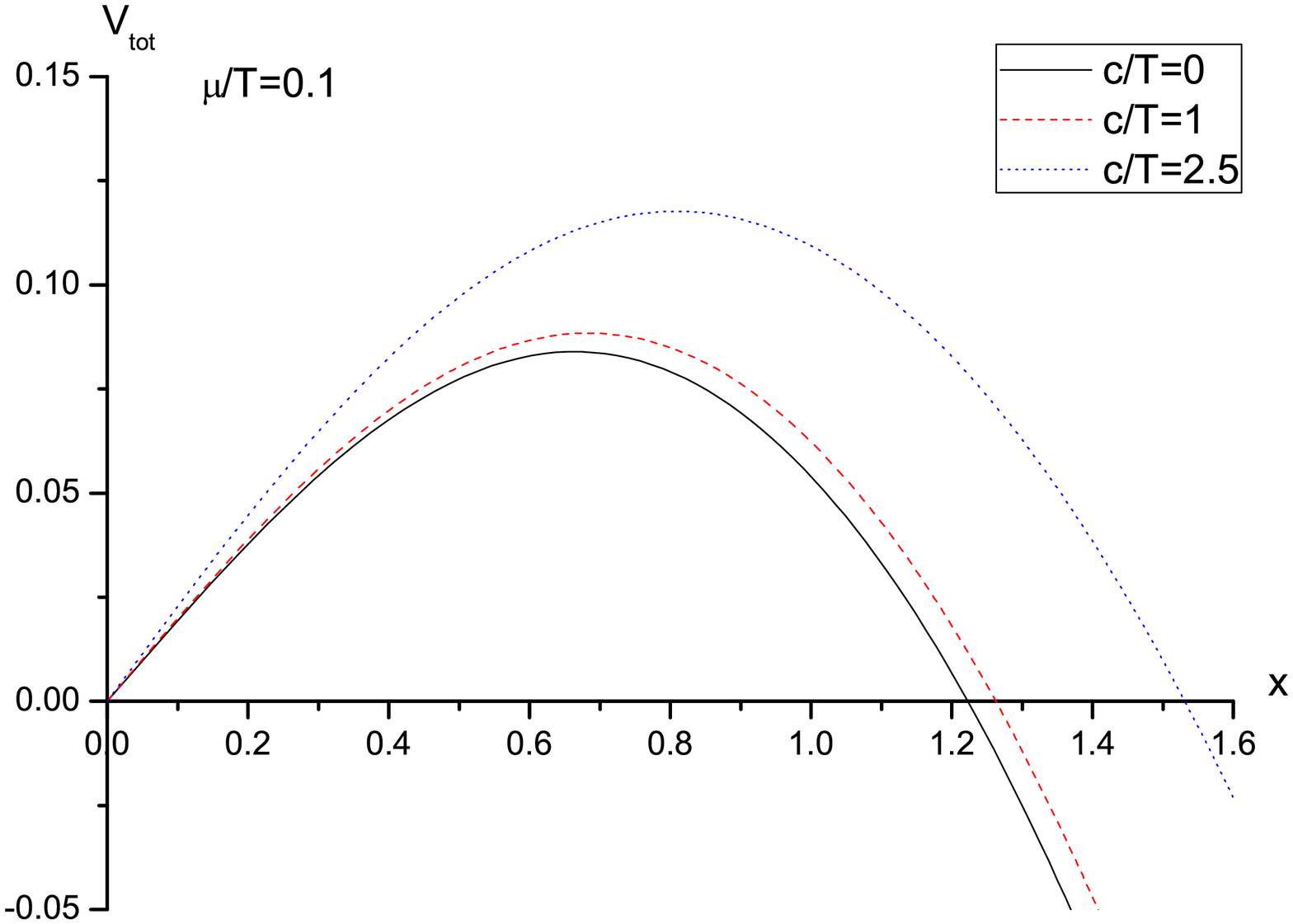}
\includegraphics[width=8cm]{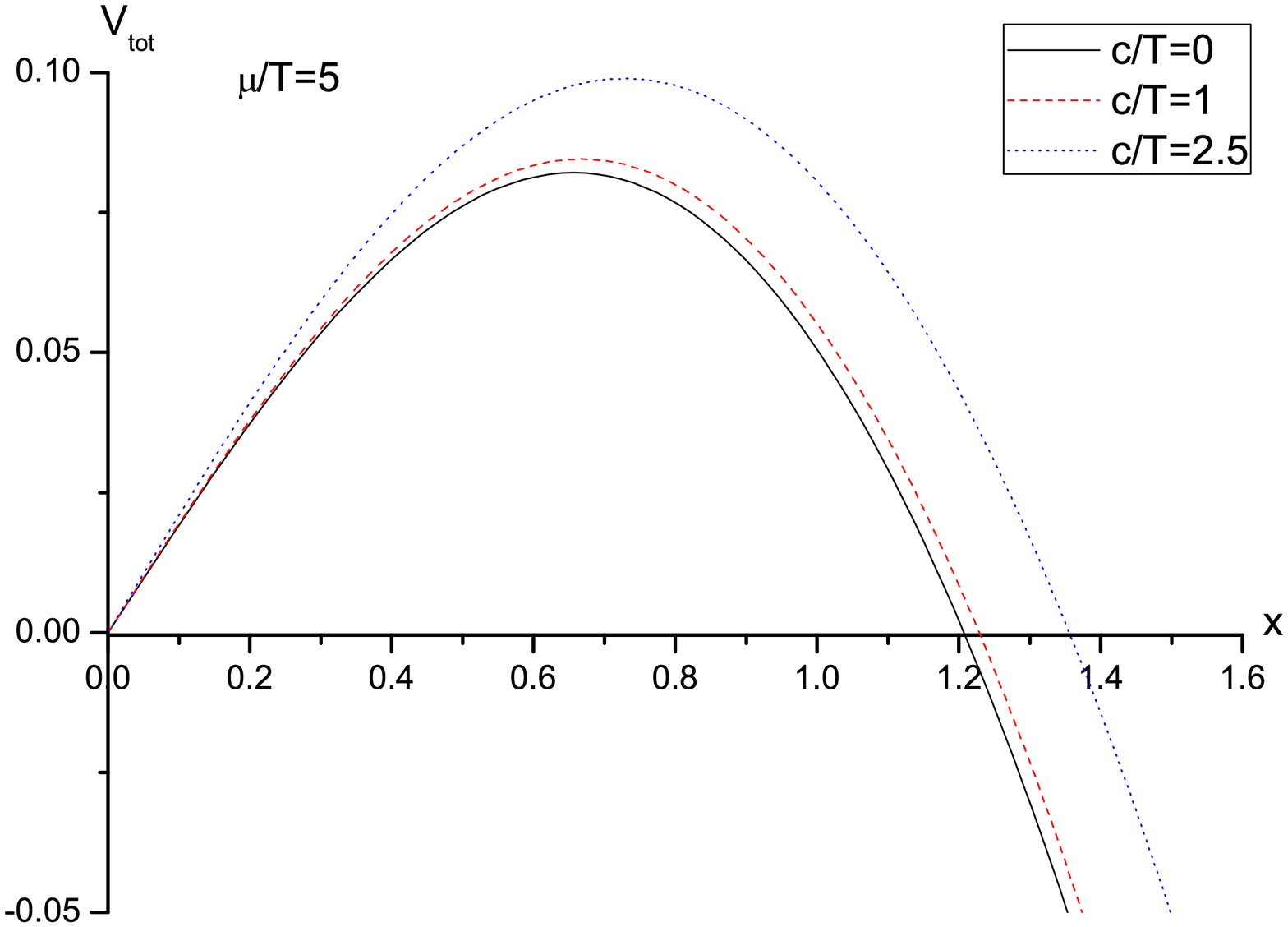}
\caption{$V_{tot}(x)$ versus $x$ with $\alpha=0.8$ and fixed
$\mu/T$ for different values of $c/T$. In both plots from top to
bottom $c/T=2.5, 1, 0$, respectively.}
\end{figure}

\begin{figure}
\centering
\includegraphics[width=8cm]{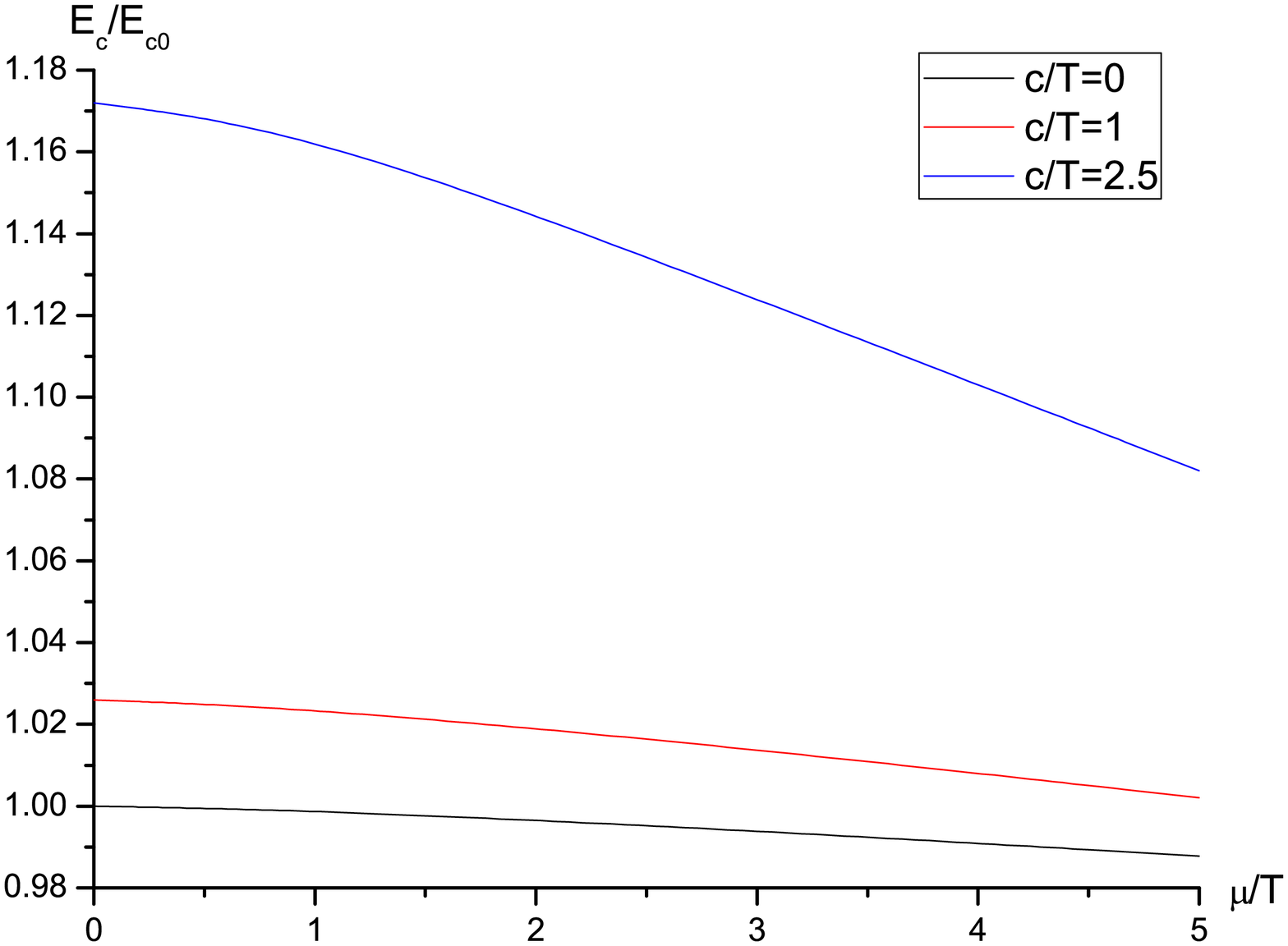}
\includegraphics[width=8cm]{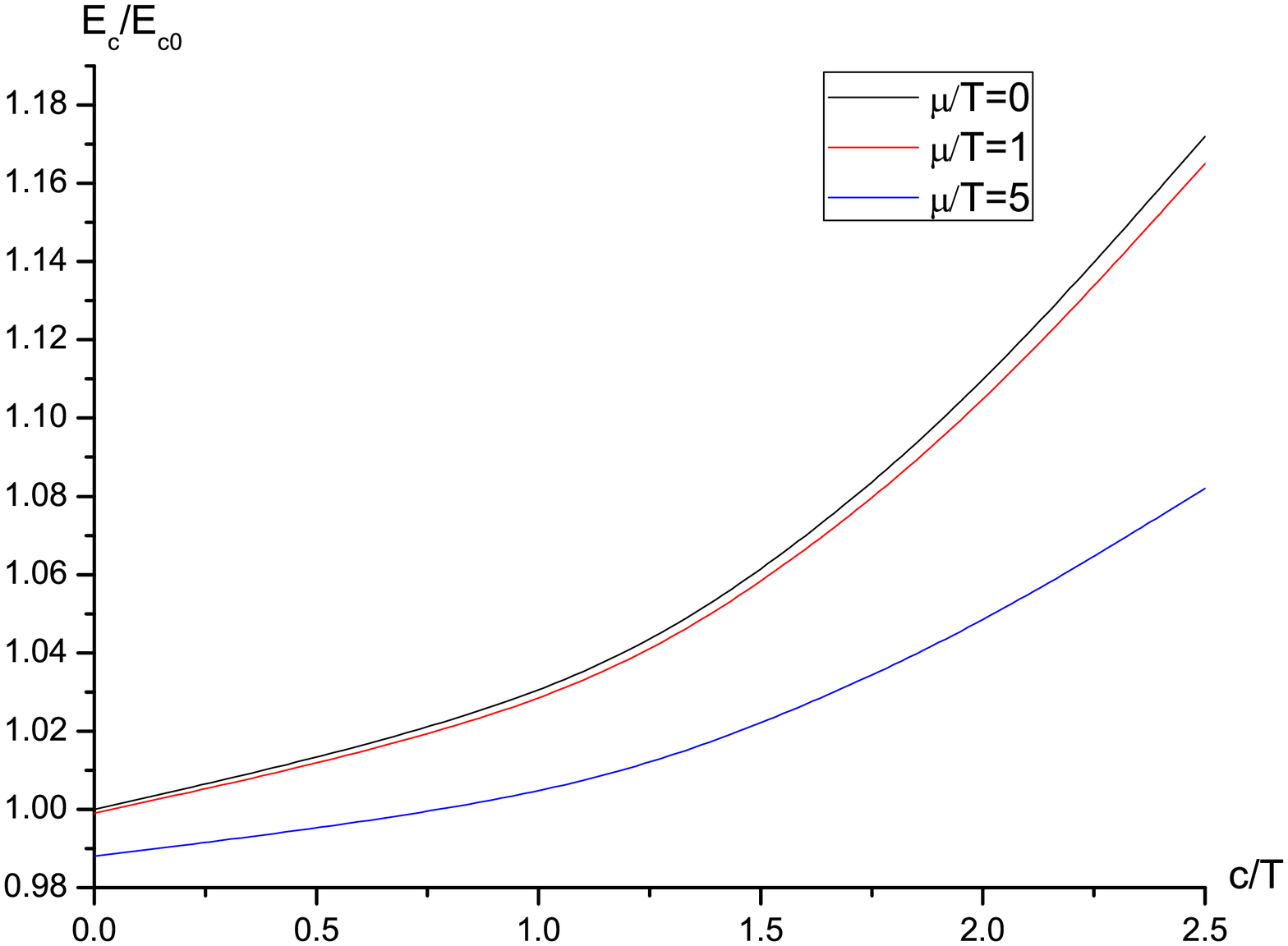}
\caption{Left: $E_c/E_{c0}$ versus $\mu/T$, from top to bottom
$c/T=2.5, 1, 0$, respectively. Right: $E_c/E_{c0}$ versus $c/T$,
from top to bottom $\mu/T=0, 1, 5$, respectively. }
\end{figure}

Also, we plot $V_{tot}(x)$ against $x$ with fixed $\mu/T$ for
different values of $c/T$ in fig.5. One finds at fixed $\mu/T$,
the height and width of the potential barrier both increase as
$c/T$ increases, implying the presence of confining scale reduces
the Schwinger effect, reverse to the effect of chemical potential.

Finally, to understand how chemical potential and confining scale
affect the critical electric field, we plot $E_c/E_{c0}$ versus
$\mu/T$ ($c/T$) in the left (right) panel of fig.6, where $E_{c0}$
denotes the critical electric field of SYM. One can see that
$E_c/E_{c0}$ decreases as $\mu/T$ increases, indicating the
chemical potential decreases $E_c$ thus enhancing the Schwinger
effect. Meanwhile, the confining scale has an opposite effect,
consistently with the potential analysis. Furthermore, it can be
seen that $E_c/E_{c0}$ can be larger or smaller than one, which
means that the SW$_{T,\mu}$ model may provide a wider range of the
Schwinger effect in comparison to SYM.

\section{conclusion}
The study of Schwinger effect in non-conformal plasma under the
influence of chemical potential may shed some light on heavy ion
collisions. In this paper, we investigated the effect of chemical
potential and confining scale on the holographic Schwinger effect
in a soft wall AdS/QCD model. We analyzed the electrostatic
potentials by evaluating the classical action of a string
attaching the rectangular Wilson loop on a probe D3 brane sitting
at an intermediate position in the bulk AdS and calculated the
critical electric field from DBI action. We found that the
inclusion of chemical potential tends to decrease the potential
barrier thus enhancing the production rate, reverse to the effect
of confining scale. Moreover, we observed with some chosen values
of $\mu/T$ and $c/T$, $E_c$ can be larger or smaller than it
counterpart of SYM, implying the SW$_{T,\mu}$ model may provide
theoretically a wider range of the Schwinger effect in comparison
to SYM.

However, there are some questions need to be studied further.
First, the potential analysis are basically within the Coulomb
branch, related to the leading exponent corresponding to the
on-shell action of the instanton, not the full decay rate. Also,
the SW$_{T,\mu}$ model is not a consistent model since it does not
solve the full set of equations of motion. Performing such
analysis in some consistent models, e.g.
\cite{JP,AST,DL,DL1,SH,SH1,RRO} would be instructive (usually the
metrics of those models are only known numerically, so the
calculations are more challenging).

\section{Acknowledgments}
This work is supported by the NSFC under Grant No. 11705166 and
the Fundamental Research Funds for the Central Universities, China
University of Geosciences (Wuhan) (No. CUGL180402).


\end{document}